\documentclass[manuscript,screen,review=False]{acmart}
\usepackage{wasysym}
\usepackage{subfiles}
\usepackage[show]{chato-notes}
\usepackage{multicol}
\usepackage{circledsteps}
\usepackage{booktabs}
\usepackage{hhline}

\usepackage{multirow}

\AtBeginDocument{%
  \providecommand\BibTeX{{%
    \normalfont B\kern-0.5em{\scshape i\kern-0.25em b}\kern-0.8em\TeX}}}

\usepackage{xspace}
\newcommand{\ModelName}{Sheaf4Rec\xspace}

\begin{document}

\title{\ModelName: Sheaf Neural Networks for Graph-based Recommender Systems}

%\authornote{Both authors contributed equally to %this research.}
%\email{webmaster@marysville-ohio.com}
%\affiliation{%
%  \institution{Institute for Clarity in Documentation}
%  \streetaddress{P.O. Box 1212}
%  \country{USA}

\author{Antonio Purificato}
\affiliation{%
  \institution{Sapienza University of Rome}
  \city{Rome}
  \country{Italy}}
\email{purificato@diag.uniroma1.it}

\author{Giulia Cassarà}
\affiliation{%
  \institution{Sapienza University of Rome}
  \city{Rome}
  \country{Italy}}
\email{giulia.cassara@uniroma1.it}

\author{Federico Siciliano}
\affiliation{%
  \institution{Sapienza University of Rome}
  \city{Rome}
  \country{Italy}}
\email{siciliano@diag.uniroma1.it}

\author{Pietro Liò}
\affiliation{%
  \institution{University of Cambridge}
  \city{Cambridge}
  \country{United Kingdom}}
\email{pietro.lio@cl.cam.ac.uk}

\author{Fabrizio Silvestri}
\affiliation{%
  \institution{Sapienza University of Rome}
  \city{Rome}
  \country{Italy}}
\email{fsilvestri@diag.uniroma1.it}

\renewcommand{\shortauthors}{Purificato et al.}

\begin{abstract}
%Current state & limitations
Recent advancements in Graph Neural Networks (GNN) have facilitated their widespread adoption in various applications, including recommendation systems. GNNs have proven to be effective in addressing the challenges posed by recommendation systems by efficiently modeling graphs in which nodes represent users or items and edges denote preference relationships. However, current GNN techniques represent nodes by means of a single static vector, which may inadequately capture the intricate complexities of users and items.
%Solution
To overcome these limitations, we propose a solution integrating a cutting-edge model inspired by category theory: \ModelName. Unlike single vector representations, Sheaf Neural Networks and their corresponding Laplacians represent each node (and edge) using a vector space. Our approach takes advantage from this theory and results in a more comprehensive representation that can be effectively exploited during inference, providing a versatile method applicable to a wide range of graph-related tasks and demonstrating unparalleled performance.
%Numbers
Our proposed model exhibits a noteworthy relative improvement of up to 8.53\% on F1-Score@10 and an impressive increase of up to 11.29\% on NDCG@10, outperforming existing state-of-the-art models such as Neural Graph Collaborative Filtering (NGCF), KGTORe and other recently developed GNN-based models.
In addition to its superior predictive capabilities, \ModelName shows remarkable improvements in terms of efficiency: we observe substantial runtime improvements ranging from 2.5\% up to 37\% when compared to other GNN-based competitor models, indicating a more efficient way of handling information while achieving better performance.
Code is available at \url{https://github.com/antoniopurificato/Sheaf4Rec}.

% Our SNN approach outperforms traditional baseline techniques on F1-Score@10, achieving improvements of 5.8\% on MovieLens 1M, and 2.8\% in terms of Recall@100 on Book-Crossing for collaborative filtering. %Additionally, it demonstrates a 1.6\% enhancement in terms of Hits@20 on the OGB Large-Scale Challenge for the DDI dataset in link prediction tasks.
% In our SNN approach, we sensitively outperform two state-of-the-art recommendation models; we observe a relative improvement of F1@10on Neural Graph Collaborative Filtering (NGCF) by 2.4\% and on LightGCN by 1.6\% under identical experimental conditions on the MovieLens 100k dataset.

%\noindent Code is available at \url{https://anonymous.4open.science/r/SheafNNS_Recommender_System-5886}.
\end{abstract}




%% A "teaser" image appears between the author and the affiliation


\maketitle
\section{Introduction}
% Current state: GNNs
Graph Neural Networks (GNNs) \cite{4700287} have demonstrated remarkable capabilities, achieving outstanding performance in a variety of tasks, with notable examples including Applied Calculus \cite{https://doi.org/10.48550/arxiv.2206.00711}, Drug discovery \cite{duvenaud2015convolutional}, and Natural Language Processing \cite{marcheggiani-titov-2017-encoding}. GNNs excel especially in contexts where relationships play a central role, such as Internet information services, which by their nature represent data in relational graphs structures \cite{sankar2021graph}. For instance, social media relationships can be modelled as a unified graph, with nodes representing individuals and edges representing mutual connections.

% Current state: GNNs in Recommendation
Expanding on the idea of using graphs to represent complex relationships, GNNs find a standout real-world application in collaborative filtering for recommender systems \cite{herlocker2000explaining}. Collaborative filtering methods work by predicting user ratings for items based on the decisions of users with similar preferences. This approach is central to many contemporary recommender systems, seen both in operational environments \cite{linden2003amazon} and at the forefront of research \cite{https://doi.org/10.48550/arxiv.1708.05031}. The relations between users and items can in fact be visualized as a bipartite user-item graph, where labeled edges denote observed interactions. With the rise of Deep Learning techniques, including GNNs, recommendation models have reached unparalleled benchmarks \cite{https://doi.org/10.48550/arxiv.1605.09477, wu2022graph}. This reflects a transformative trend in the use of advanced graph-based models to improve the performance of recommender systems.

% Problem
Despite their impressive performance, Graph Neural Networks (GNNs) have limitations in representing the complex interactions between users and items \cite{alon2021bottleneck}. The conventional approach of encapsulating these interactions as a single vector proves insufficient in capturing the intrinsic characteristics of the relationships. 

%``\textit{raw}
%Solution
A more comprehensive solution emerges with the introduction of a novel class of GNN architectures known as ``Sheaf Neural Networks'' \cite{https://doi.org/10.48550/arxiv.2202.04579}. Sheaf Neural Networks (SNNs) represent a novel class of Graph Neural Network models, drawing inspiration from Category Theory \cite{https://doi.org/10.48550/arxiv.2202.04579}.
Unlike traditional GNNs, SNNs exploit an entire vector space, allowing the encapsulation of all aspects of users or items. The vectors within this space are only actualised at inference time, allowing for a dynamic and nuanced representation that goes beyond the limitations of static vectors.
SNNs have been applied to various tasks and have demonstrated superiority over traditional GNNs \cite{duta2023sheaf}.

% Contributions
Our approach, \ModelName, links the theoretical insights of SNNs to the practical domain of item recommender systems. Specifically, our work shows
that the use of SNNs in graph-based recommender systems significantly improves their performance.

The primary original contributions of our research are as follows:
\begin{itemize}
    \item \textbf{Novel architecture for recommender system:}
    We introduce \ModelName,  a novel architecture for recommender systems utilizing Sheaf Neural Networks. This approach consistently achieves state-of-the-art performance in its designated tasks, highlighting not only the rigorous mathematical formalism of SNNs, but also their practical applicability in scenarios where relationship information is essential for constructing holistic representations of the objects involved. This is especially relevant when the inherent characteristics of these objects are ambiguous when considered in isolation.

    \item \textbf{Performance Evaluation Across Multiple Datasets:}
    Our extensive experimentation across multiple datasets for top-K recommendation underscores the viability and promise of our solution. Notably, \ModelName surpasses all existing benchmarks in the focal tasks of this study. For instance, we achieve relative improvements over state-of-the-art methods in terms of NDCG@10 by 11.29\% on the Facebook Books dataset, by 8.77\% on the Yahoo! Movies dataset in terms of F1@20 and by 8.24\% on the MovieLens 1M dataset in terms of NDCG@20.
    
    \item \textbf{Efficiency gains in recommendation computation:}
    Leveraging the advancements of our sheaf-based architecture, recommendation computations are significantly expedited. Across the various datasets used in our studies, \ModelName shows relative efficiency improvements ranging from 2.5\% to 37\%, underscoring the practical efficiency gains that can be achieved by adopting our proposed SNN-based approach in item recommendation systems.

    % \item \textbf{Relationship between model and loss function:}
    % Through a series of experiments, we elucidate the relationship between our model and the application of the Bayesian Personalised Ranking loss function.
\end{itemize}

%Experimental results indicate that our SNN-centric models surpass all existing state-of-the-art models in their designated tasks. Notably, for recommendation tasks, our approach outdoes the leading GAT method \cite{https://doi.org/10.48550/arxiv.1710.10903} by relative margins of 8.8\% and 7.3\% on the F1@10 metric for MovieLens 100k and MovieLens 1M, respectively. Additionally, we achieve a relative improvement of 8.5\% in terms of Recall@100 for Book-Crossing. Our SNN model exhibits marked improvements over the latest three GNN-based models when comparing run-time efficiencies. Specifically, our method registers relative gains ranging from 1.5\% to 27\% compared to the second most efficient technique, GAT \cite{https://doi.org/10.48550/arxiv.1710.10903}, across the datasets used in our evaluations.
% In the case of link prediction, we have a 1.6\% improvement on the ogbl-ddi dataset.
%On the MovieLens 1M dataset, SNN exhibits more than 7.3\% relative improvement on F1@10, and 2.5\% relative improvement on F1@100, over the Graph Attention Network (GAT) baseline. On the Book-Crossing dataset, SNN has a relative improvement on Recall@100 of 6.8\% in terms of Recall@10, and a relative improvement of 8.5\% in terms of Recall@100.

The rest of the paper is organized as follows. Section \ref{related_work} provides a comprehensive review of related works. In Section \ref{background} we offer an overview of Graph Neural Networks and sheaf theory, laying the groundwork for our approach. Our methodology is delineated in Section \ref{method}, followed by a discussion on implementation details, encompassing dataset selection and experimental configurations, in Section \ref{experiments}. Section \ref{results} presents a thorough analysis of the efficacy of our proposed solution, while Section \ref{conclusions} encapsulates our findings and offers insights into potential avenues for future research.

\section{Related work}
\label{related_work}
%In this section, we review the related work in the area of recommender systems.
Recommender systems have been extensively studied, and various approaches have been proposed in the literature.
%We focus on the key developments and the most relevant work in the following areas: Autoencoder-based recommender systems, matrix factorization-based, and GNNs-based approaches.

\subsection{Matrix Factorization-Based Recommender Systems}
Matrix Factorization (MF) is a popular collaborative filtering technique that aims to decompose a user-item interaction matrix into lower-dimensional latent factors. This low-rank approximation can capture the underlying structure of the data and yield better recommendations \cite{bokde2015matrix}.

One of the most well-known MF-based methods for recommender systems is Singular Value Decomposition (SVD) \cite{ZHOU2015717}. However, traditional SVD struggles to deal with missing data, which is common in recommendation scenarios \cite{koren2009}. To address this issue, \citet{koren2009} introduced \emph{SVD++}, an extension of the basic SVD that incorporates implicit feedback into the factorization process.

Non-negative Matrix Factorization (NMF), introduced by Lee and Seung \cite{Lee1999-jh}, represents another significant development in MF-based recommender systems. NMF imposes non-negativity constraints on the factor matrices, which can aid in interpretability and improve recommendation performance.

The idea of TrustMF \cite{10.1109/TPAMI.2016.2605085} was to improve recommendation by using more data. In addition to explicit rating data, TrustMF also leverages the social trust relationships between users as auxiliary information. Trust is a crucial social cue because people are more inclined to accept the opinions of those they trust.
By factorizing the trust network according to the directional feature of trust, the users are mapped to two low-dimensional spaces: trustee and truster. Next, separate models of the truster and trustee space are created to extract latent features of individuals and objects.

% Factorization Machines (FM) \cite{Rendle2013-pk} were introduced as a generalization of Matrix Factorization (MF) capable of modeling interactions in sparse data and accommodating side information.

\subsection{Autoencoder-based recommender systems}
An inherent limitation of MF-based recommender systems is their linear nature, which constrains their ability to capture complex relationships between users and items \cite{10.1145/1921591.1921593}. Additionally, these systems generally disregard any side information (e.g., user demographics or item attributes) that could potentially benefit the recommendation process. Furthermore, MF methods can struggle in cold-start situations where little or no information about new users or items is available \cite{10.1145/1921591.1921593}.
To address these limitations, Autoencoder-based recommender systems have emerged as a promising solution. Autoencoders are unsupervised learning models that aim to learn a compressed representation of input data by encoding it into a lower-dimensional space and then reconstructing the original input from the encoded representation \cite{Zhang_2019}.

One of the early works in autoencoder-based recommender systems was presented by \citet{Sedhain2015-gn}, who proposed AutoRec, an autoencoder framework for CF. AutoRec learns a latent feature representation of users or items, capturing the nonlinear relationships between them.
Building on this, \citet{Liang2015-id} introduced the Variational Autoencoder (VAE) for CF, which learns a probabilistic latent representation to better model the uncertainty in user-item preferences.

Recommendation via Dual-Autoencoder (ReDa) \cite{zhuang2017representation} takes a novel approach by using two AEs to simultaneously learn hidden latent representations for users and items, minimizing the deviations in the training data.
However, ReDa considers only explicit feedback information between users and items, such as rating matrix and check-in matrix, and ignores side information about users or items.

\citet{https://doi.org/10.48550/arxiv.1811.09975} implement a recurrent version of the VAE, where instead of passing a subset of the whole history regardless of temporal dependencies, they pass the consumption sequence subset through a recurrent neural network.

\citet{9224132} propose EnsVAE, which is based on mapping the original data to higher-order features
interactions. EnsVAE  learns the interest probability distribution on a per-item basis, allowing the variational autoencoder to correctly learn and represent the distribution pattern. Moreover, by sampling from the distribution, complete rows of new users and columns of new resources are generated even with minimal interaction.

%Han \emph{et al.} \cite{Han_2021} improve the previous idea. They first pre-train an autoencoder with the local kernelised weight matrix, which transforms the data from one space into the feature space by using a 2d-RBF kernel. Then, the pre-trained autoencoder is fine-tuned with the rating matrix, produced by a convolution-based global kernel, which captures the
%characteristics of each item.

\subsection{Graph Neural Networks-based Recommender Systems}
When dealing with sparse datasets, autoencoders struggle to accurately reconstruct the input \cite{kuchaiev2017training}, resulting in latent representations that fail to adequately generalize to unseen data or capture the nuances of user preferences, ultimately leading to poor recommendations.

In contrast, Graph Neural Networks (GNNs) have emerged as a powerful class of deep learning models, gaining attention due to their ability to handle complex relational data. Their application in recommender systems has become increasingly popular, capitalising on the inherent graph structure of user-item relationships.

An early work in GNN-based recommender systems by \citet{Van_den_Berg2017-eq} introduced the Graph Convolutional Matrix Completion (GCMC) method for link prediction in bipartite graphs. GCMC applies graph convolution operations to user-item interactions, effectively learning latent representations for recommendation, showing improved performance compared to traditional matrix factorization-based methods.
%that learn latent factors for users and items by factorizing the user-item interaction matrix.

NGCF \cite{Wang2019-gl} designs an innovative embedding propagation layer that refines user or item embeddings by aggregating those of interacted items or users. The embedding refinement step explicitly injects collaborative
signal into embeddings. By stacking multiple embedding propagation layers, the model captures collaborative signals in higher order connectivities.

STAR-GCN \cite{zhang2019star} adopts a multi-block architecture featuring a stack of GCN encoder-decoders combined with intermediate supervision. This approach allows the propagation of the initial node embeddings and features across the whole graph via an encoder-decoder architecture. The graph encoder captures semantic graph structures and input content features, while the decoder works to recover the input node embeddings. In contrast to GCMC \cite{Van_den_Berg2017-eq}, utilizing one-hot encoding node inputs, STAR-GCN employs low-dimensional user and item latent factors as inputs to limit the complexity of the model space.

LightGCN \cite{He2020-pe}, based on the GCN approach, reduces the computational cost by adopting a straightforward weighted sum aggregator and discarding feature transformation and nonlinear activation. This streamlined variant reduces the trainable parameters to the embeddings in the initial layer. The model uses layer combination to avoid oversmoothing, make the final representation more comprehensive and capture the effect of graph convolution with self-connections.

In summary, GNNs have shown promising results in recommender systems by successfully modeling the complex relationships between users and items, contributing to improved recommendation quality and user satisfaction \cite{10.1145/3568022}.

\section{Background}
\label{background}

\begin{table}[ht]
  \begin{tabular}{cc}
    \toprule
    Symbol & Description\\ 
    \midrule
    $v_i$ & Node $i$\\
    $e_{ij}$ & Edge from $v_i$ to $v_j$ \\
    $N(v)$ & Neighbourhood of node $v$ \\
    $H(l)$ & Hidden state at node $l$\\
    $G$ & Graph \\
    $V$ & Set of nodes \\
    $E$ & Set of edges \\
    $A$ & Adjacency matrix\\
    $\mathcal{F}_{v \unlhd e}$ & Restriction map between node-edge pair $v \unlhd e $\\
    $C^i$ & Space of $i$-cochains \\
    $I_n$ & Identity matrix of size $n$\\

    \bottomrule
  \end{tabular}
  \caption{Notation used in the paper.}
\label{table:notation}
\end{table}

We provide a brief overview of the basic concepts that are essential for understanding our proposed solution. First, we outline the main characteristics of GNNs, followed by an introduction to cellular sheaf theory. Finally, we present the learning algorithm central to out approach: \textit{Neural Sheaf Diffusion}. The following discussion summarizes the key theoretical definitions and methods introduced by \citet{https://doi.org/10.48550/arxiv.2206.08702}.
The notation used throughout this paper is illustrated in \ref{table:notation}.

% Sheaf 
% The foundational building blocks for these models are \textit{Cellular Sheaves}. In this context, cellular sheaves allocate a vector space to every node and edge within a graph and define linear maps between these spaces. Essentially, SNNs are Graph Neural Networks that operate on \textit{Cellular Sheaves}, which are vector spaces linked to the graph's nodes. The functionality of SNNs hinges on computing a generalized version of the widely recognized Graph Laplacian, referred to as the \textit{Sheaf Laplacian}~\cite{hansen2019toward}. The Sheaf Laplacian is genuinely a generalization of the Graph Laplacian; when the vector spaces of the cellular sheaves are confined to one dimension and identity maps are established between them, the two Laplacians align perfectly. Notably, the Sheaf Laplacian is determined non-parametrically via the restriction map during the pre-processing phase \cite{https://doi.org/10.48550/arxiv.2206.08702}.

\subsection{Graph Neural Networks}
The rise of GNNs can be attributed to the advancement in both Neural Networks (NNs) and Graph Representation Learning (GRL) \cite{9046288}. NNs excel at extracting localized features from Euclidean data, such as images or texts. However, adapting NNs to non-Euclidean data, like graphs, requires generalization to accommodate variable object sizes (e.g. nodes in graphs) \cite{li2019deepgcns}. On the other hand, GRL focuses on generating low-dimensional vector representations for nodes, edges, or subgraphs of a given graph, effectively capturing the intricate connectivity structures inherent in graphs.

A directed graph $G = (V, E)$ consists of a set of nodes $V$ and a set of edges $E$. Let $v_i \in V$ be a node and $e_{ij} =(v_i,v_j)\in E$ be an edge pointing from $v_i$ to $v_j$. The set of edges can also be described by an adjacency matrix $A$ as follows:
\begin{equation}
    A_{ij} = 
    \begin{cases}
    1 \,\text{if } \,e_{ij} \in E\\
    0 \, \text{otherwise}
    \end{cases}
\end{equation}
If the graph is undirected, we have $e_{ij} = e_{ji} \in E$ and $a_{ij} = a_{ji} \in A$, thus the adjacency matrix is symmetric.

The neighborhood of a node $v$ is denoted as:
\begin{equation}
    N(v) = \{u \in V \mid (v,u) \in E\}
\end{equation}

The degree matrix $D$ of $G$ contains information about the number of edges attached to each vertex:
\begin{equation}
    D_{ij} = 
    \begin{cases}
    |N(i)| & \text{if } i=j \\
    0 & \text{otherwise}
    \end{cases}
\end{equation}
Using the definitions of degree matrix $D$ and adjacency matrix $A$, the Laplacian matrix $L$ is easily expressed as $L = D - A$.

Given graph data, the primary objective of GNNs is to iteratively aggregate feature information from neighbors and incorporate this aggregated information with the current central node representation during the propagation process. From a network architecture standpoint, GNNs consist of several propagation layers, which involve aggregation and update operations.

The update operation at the $l$-th layer is expressed as:
\begin{equation}
H^{(l)} = f(H^{(l-1)},A)
\end{equation}
Where $H^{(l)} \in \mathbb{R}^{|V| \times d}$ denotes the hidden state of nodes at layer $l$, $A$ is the adjacency matrix, and $f$ is a function that combines, for each node, the previous layer's hidden state $H^{(l-1)}$ with its neighbours' aggregated information. %(taken from A)?

% The aggregation operation at the $l$-th layer is formulated as:
% \begin{equation}
% \alpha^{(l)} = g\left(A,H^{(l-1)}\right)
% \end{equation}
% where $g$ is an activation function and $H^{(l-1)} \in \mathbb{R}^{|V| \times d}$ denotes the hidden state of nodes at layer $l-1$.

%%Aggiungere caso con D e W
%Per esempio, nel caso di GCN?, queste due operazioni possiamo scriverle come

For example, in the case of GCN \cite{https://doi.org/10.48550/arxiv.1609.02907}, the update operation can be defined as:

\begin{equation}
H^{(l)} = h\left(D^{-\frac{1}{2}}AD^{-\frac{1}{2}}H^{(l-1)}W^{(l)}\right)
\end{equation}

Where $h$ is a non-linear activation function, $D$ is the degree matrix, $A$ is the adjacency matrix, $H^{(l-1)} \in \mathbb{R}^{|V| \times d}$ denotes the hidden state of nodes at layer $l-1$ and $W^{(l)}$ is the matrix of learnable weights at layer $l$.
%where h is an activation function and D is degree... W is weights

%I_n tolto, ma diciamo che ci sono i self-loops, quindi A_{ii} = 1.

By stacking multiple layers of aggregation and update operations, GNNs can learn expressive node representations that effectively capture both local and global graph properties, making them suitable for graph-related tasks such as node classification, link prediction, and graph classification.

\subsection{Cellular Sheaf Theory}
Cellular sheaves provide a framework for analyzing graph-structured data by combining sheaf theory \cite{shepard1985cellular} with graph theory \cite{west2001introduction}.

In a cellular sheaf, data is associated with both the nodes and edges of a graph, effectively representing local and global structures. This section comprehensively explains the formulation of cellular sheaves as applied to graphs.

%We first define the data associated with the nodes and edges to construct a cellular sheaf on
Given an undirected graph $G = (V, E)$, for each node $v \in V$, we associate a vector space or algebraic structure $\mathcal{F}(v)$. Similarly, for each edge $e \in E$, we associate a vector space or algebraic structure, $\mathcal{F}(e)$. $\mathcal{F}(v)$ and $\mathcal{F}(e)$ the local data on the graph.

A restriction map $\mathcal{F}_{v \unlhd e}: \mathcal{F}(v) \rightarrow \mathcal{F}(e)$ represents the relationship between local data on nodes and edges for each incident node-edge pair $v \unlhd e$. In this context, the vector spaces $\mathcal{F}(v)$ and $\mathcal{F}(e)$, corresponding to nodes and edges, are called stalks.
%while the linear maps that connect them are known as restriction maps \cite{https://doi.org/10.48550/arxiv.2206.08702}.
The restriction maps play a key role in transforming cellular sheaves into a tangible representation that can be used to generate recommendations.

%WE take the notation from \cite{https://doi.org/10.48550/arxiv.2202.04579}
%We use the same notation as \cite{https://doi.org/10.48550/arxiv.2202.04579}

For a given sheaf $(G,\mathcal{F})$, the space of $0$-cochains $C^0(G,\mathcal{F})$ is defined as the direct sum over the vertex stalks:
\begin{equation}
    C^0(G,\mathcal{F}) = \bigoplus_{v \in V}\mathcal{F}(v)
\end{equation}
Similarly, the space of $1$-cochains $C^1(G,\mathcal{F})$ encompasses the direct sum over the edge stalks.

By assigning an arbitrary orientation to each edge $e$, the co-boundary map serves as a linear map associating $0$-cochains to $1$-cochains by capturing the difference between the data associated with the vertices connected by an edge. Formally, the co-boundary map is defined as:
\begin{equation}
    \delta : C^0(G,\mathcal{F}) \rightarrow C^1(G,\mathcal{F}) = \delta(x)_e = \mathcal{F}_{v \unlhd e}x_v - \mathcal{F}_{u \unlhd e}x_u
\end{equation}

The sheaf Laplacian is an operator that generalizes the concept of the graph Laplacian to the setting of sheaves on graphs \cite{hansen2019toward}. The graph Laplacian is a matrix that captures the connectivity structure of a graph, and it is widely used in graph analysis and spectral graph theory \cite{Fiedler1973}. The sheaf Laplacian extends these ideas to capture not only the graph's connectivity but also the local-to-global relationships of the data on the graph as described by the sheaf.
The sheaf Laplacian is an operator that maps $0$-cochains to $0$-cochains:
\begin{equation}
    L_{\mathcal{F}} = \delta^T\delta = \sum_{v,u \unlhd e} \mathcal{F}_{v \unlhd e}^T(\mathcal{F}_{v \unlhd e} x_v - \mathcal{F}_{u \unlhd e}x_u)
\end{equation}
Notably, the sheaf Laplacian is closely related to the Laplacian, but it takes restriction maps into account instead of edges.
The sheaf Laplacian captures both the graph's connectivity structure and the local-to-global relationships of the data described by the sheaf.

Finally, the normalized sheaf Laplacian $\Delta_{\mathcal{F}}$ is expressed as:
\begin{equation}
    \Delta_{\mathcal{F}} = D^{-\frac{1}{2}}L_{\mathcal{F}}D^{-\frac{1}{2}}
\end{equation}
Here, $D$ represents the diagonal of $L_{\mathcal{F}}$.

%block-diagonal of $L_{\mathcal{F}}$, i.e. a matrix where all non-zero elements are arranged in square blocks along the main diagonal, and all off-diagonal blocks are zero matrices.

\subsection{Neural Sheaf Diffusion}
Considering a graph $G = (V, E)$, each individual node $v \in V$ is associated with a $d$-dimensional feature vector $x_v \in \mathcal{F}(v)$. The individual vectors $x_v$ are column-stacked to create an $nd$-dimensional vector $x \in C^0(G, \mathcal{F})$. The vectors belonging to $C^0(G, \mathcal{F})$ form the columns of the feature matrix $X \in \mathbb{R}^{(nd) \times f}$.

Sheaf diffusion can then be described as a process that operates on $(G, \mathcal{F})$, controlled, at time $t$, by the differential equation:
\begin{equation}
    X(0) = X,\quad \dot{X}(t) = -\Delta_{\mathcal{F}}X(t)
\end{equation}

This equation is discretized using the explicit Euler scheme, which employs a unit step size:
\begin{equation}
    X(t+1) = X(t) - \Delta_{\mathcal{F}}X(t) = (I_{nd} - \Delta_{\mathcal{F}})X(t)
\end{equation}

In the model proposed in \cite{https://doi.org/10.48550/arxiv.2202.04579}, the discretization of the above equation is carried out as follows:
\begin{equation}
\label{averaging}
    X(t+1) = X(t) - \sigma(\Delta_{\mathcal{F}(t)}(I_n \otimes W_{1}^{t})X_tW_2^{t})
\end{equation}
In this case, the sheaf $\mathcal{F}(t)$ and the weights $W_1^{t} \in \mathbb{R}^{d \times d}$ and $W_2^{t} \in \mathbb{R}^{f_1 \times f_2}$ are time-dependent. This implies that the underlying geometric structure of the graph changes over time.
\section{Method}
\label{method}

% \begin{figure}[t]
%   \centering
%   \includegraphics[width=\textwidth]{figure/complete_model.pdf}
%   \caption{Complete representation of our approach. Initially, are computed the input embeddings related to positive and negative samples. The output of these embeddings is taken as input from \ModelName (taken from \cite{https://doi.org/10.48550/arxiv.2202.04579}, and finally, the predictions of \ModelName are used to compute the performance metrics.}
%   \label{complete_model}
% \end{figure}

%%Introduzione
We combine cellular sheaves with the Bayesian Personalized Ranking (BPR) loss to implement our novel recommender system framework. By harnessing the sheaf structure, our objective is to compute latent factors, capitalizing on the rationale inherent to the BPR loss optimization. Our methodology introduces a fresh viewpoint on personalized ranking within recommender systems by merging the robust expressiveness of the cellular sheaf framework with the established efficacy of the BPR loss.

\subsection{Data}

The data consists of a set of users $U = \mathbb{N}^n$, a set of items $I = \mathbb{N}^m$ and a set of reviews $R = \{(i,j,r_{i,j}) | i \in U, j \in I, r_{i,j} \in \mathbb{R}\}$. Tipically, the cardinality of $R$ is significantly smaller than its possible maximum, i.e. $|R| << n \cdot m$, indicating that it captures only a subset of all possible interactions between users and items.

Given this data, we transform it into a  bipartite graph $G = (V, E)$, where $V = U \cup I$ represents the disjoint union of the user vertex set $U$ and the item vertex set $I$. The set of edges $E$ containes edges connecting users to the items they have interacted with, such that if $(i,j) \in E$, then $(i,j,r_{i,j}) \in R$.

However, applying the sheaf framework directly to bipartite graphs poses challenges, as discussed in \cite{https://doi.org/10.48550/arxiv.2206.08702}. Therefore, we adapt the sheaf Laplacian operator to fit the structure of bipartite graphs. This adaptation involves computing a projection on the bipartite graph, trasforming it into a structure that is not strictly bipartite but retains the same properties. This approach preserves the connection between data points and the intrinsic properties of the sheaf, while facilitating parameter updates.

%%EMBEDDING
\subsection{Embedding}
\begin{figure}[t]
  \centering
  \includegraphics[width=\textwidth]{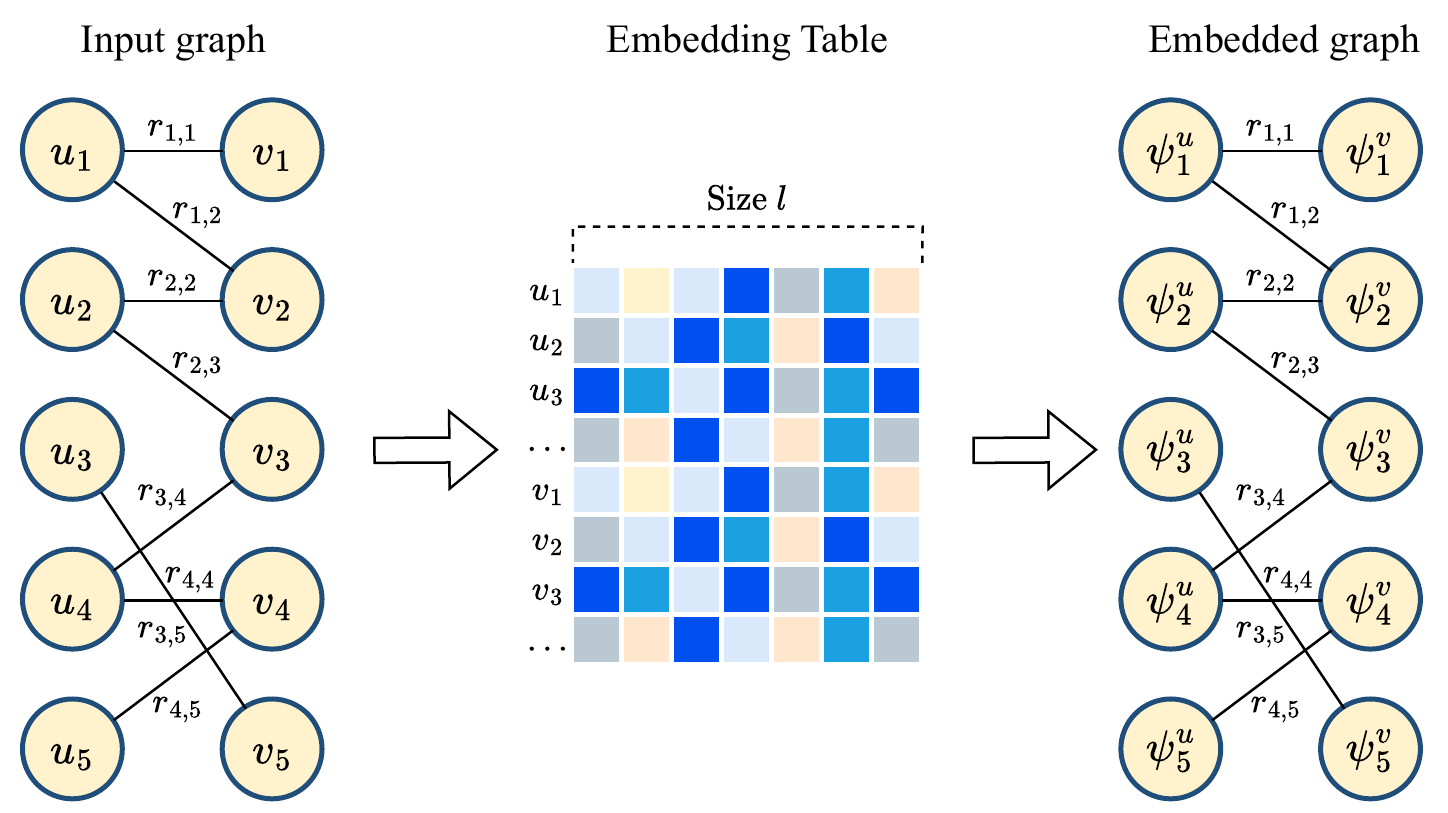}
  \caption{The input graph comprises users and items, forming a bipartite structure, with review values on edges. Using two embedding tables, one for the users and one for the items, the graph is transformed to its embedded version.}
  \label{fig:model_emb}
\end{figure}

Since nodes are initially represented only by identifiers, which lack substantial information, it becomes necessary to map them into a more meaningful latent space where they can convey richer information. This process involves defining two embedding tables, $\Psi_u$ and $\Psi_v$, which contain the embeddings for all users and items, respectively. These tables are formed as unions of individual embeddings: $\Psi_u = \bigcup_{j \in U} \psi_j^u$ and $\Psi_v = \bigcup_{i \in I} \psi_i^v$.
To illustrate, consider the transformation of user $u_1$ into its embedding $\psi_1^u$ and item $v_1$ into its embedding $\psi_1^v$.

At first, random embeddings are generated for each user and item in the graph to initialize the embedding tables. These are trained along with the model, adapting and refining their representations as the learning process progresses.

This process is visually illustrated in Figure \ref{fig:model_emb}, which shows the transformation of the bipartite graph into its embedded version.

\subsection{Sheaf Diffusion}
\begin{figure}[t]
  \centering
  \includegraphics[width=\textwidth]{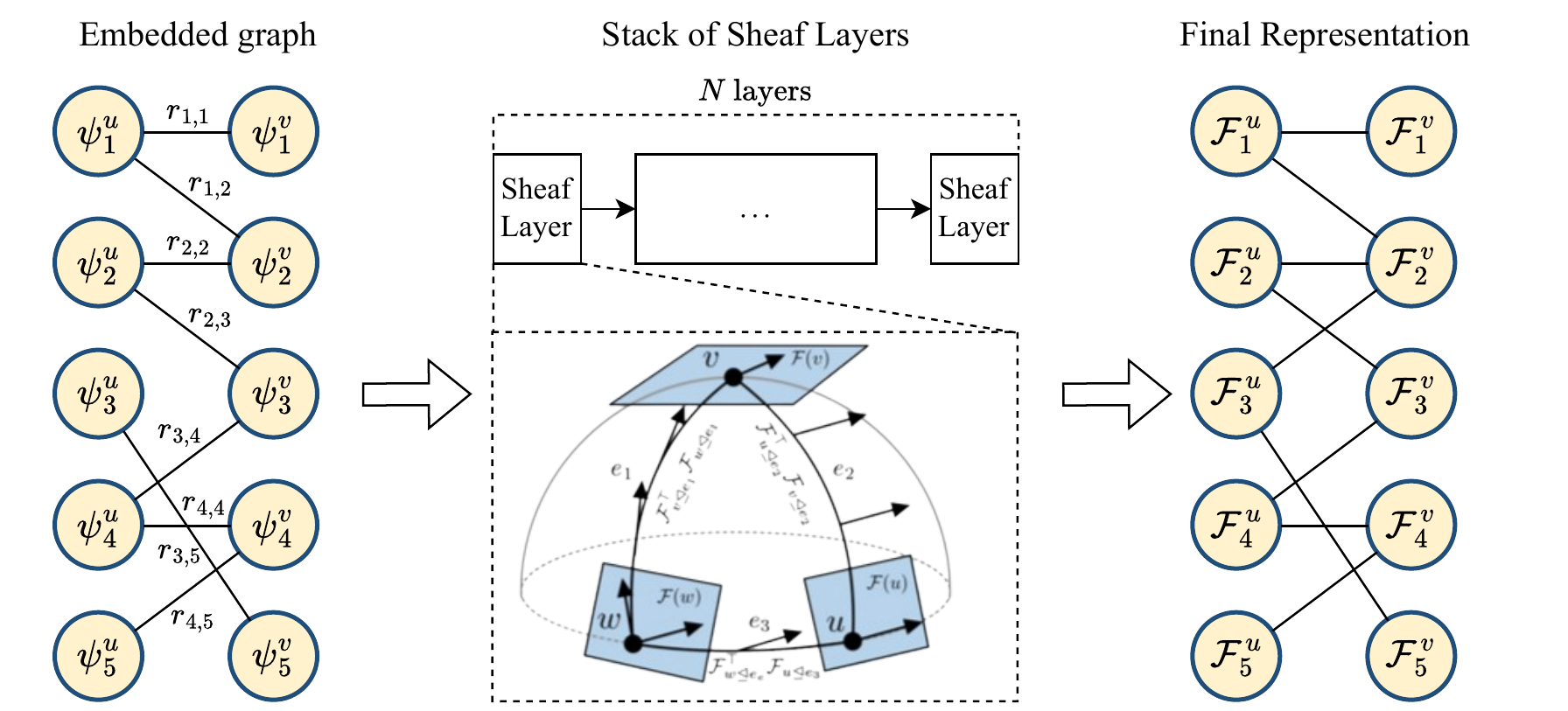}
  \caption{The input to the sheaf layers is the embedded graph obtained from the two embedding tables. In the sheaf layers, $0$-cochains are mapped to user embeddings and $1$-cochains to item embeddings. These embeddings are continuously updated. To enhance the representative power of the proposed formalism, multiple sheaf layers are stacked, resulting in better recommendations. The output of the sheaf layers is still a graph, but containing the final representation.}
  \label{fig:model_sheaf}
\end{figure}

Our novel recommender system, \ModelName, utilize cellular sheaves as the foundational framework. The cellular sheaf provides a structured representation of the underlying problem setting, allowing to effectively organize the intricate relationships between users, items, and preference scores. 

In our architecture, $0$-cochains naturally correspond to the embeddings of users $\Psi_u$ and items $\Psi_v$, while $1$-cochains naturally represent the observed preference scores $r_{i,j}$ for $i \in U$ and $j \in I$.

The input to the first sheaf layer is the embedded graph obtained from the two embedding tables. The sheaf layer aggregates each node embedding $\psi_i$ with the semantic information related to its connected nodes.
The node embedding is then updated with this new aggregated embedding.
This updating process is repeated in each sheaf layer in the network, resulting in a network composed by $N$ layers.
Through this approach, we aim to cluster users with similar item interests.

As a result, the user embeddings $\mathcal{F}_u$ in the final layer $N$ encapsulate knowledge about users and items that are up to $N$-hops away in the graph. In the context of recommendations, this process increases the embedding similarity between users with shared item interests.
A similar principle applies to the item embeddings $\mathcal{F}_v$, which represents the learned embeddings associated with each item $v$.

%The rationale behind the choice of node embedding is clear-cut: an embedding vector is generated for each node in the graph. This embedding vector can capture the graph representation and structure. 
% The decision to use node embeddings is evident: a unique embedding vector is produced for each node within the graph. These vectors are designed to encapsulate the representation and structural characteristics of the graph.
%At each step, the user's node embedding becomes more similar to other users who share the same item interests. Essentially, nodes in close proximity should also have vectors in close proximity to each other.

An inherent advantage of our approach lies in the seamless integration of the proposed embedding and the sheaf architecture. \ModelName accepts an embedded graph as input, which contains a modified representation of users and items.
%, along with the size of the graph (the sum of the number of users and items).
This cohesive integration ensures that all the pertinent information from the embeddings is stored in the corresponding vector spaces.

The sheaf diffusion process is shown in Figure \ref{fig:model_sheaf}.

%\begin{figure}[ht]
 % \centering
 % \includegraphics[width=0.6\linewidth]{figure/scheme.png}
 % \caption{The structure of our architecture. Only the information related to the user embedding  \Circled{A} exists in the input layer. In Layer 1, this information gets updated with a new item embedding  \Circled{B}. In Layer 2, the previous item embedding is updated with a new one  \Circled{C}.}
  %\label{graph}
%\end{figure}

\subsection{Recommendation}

\begin{figure}[t]
  \centering
  \includegraphics[width=\textwidth]{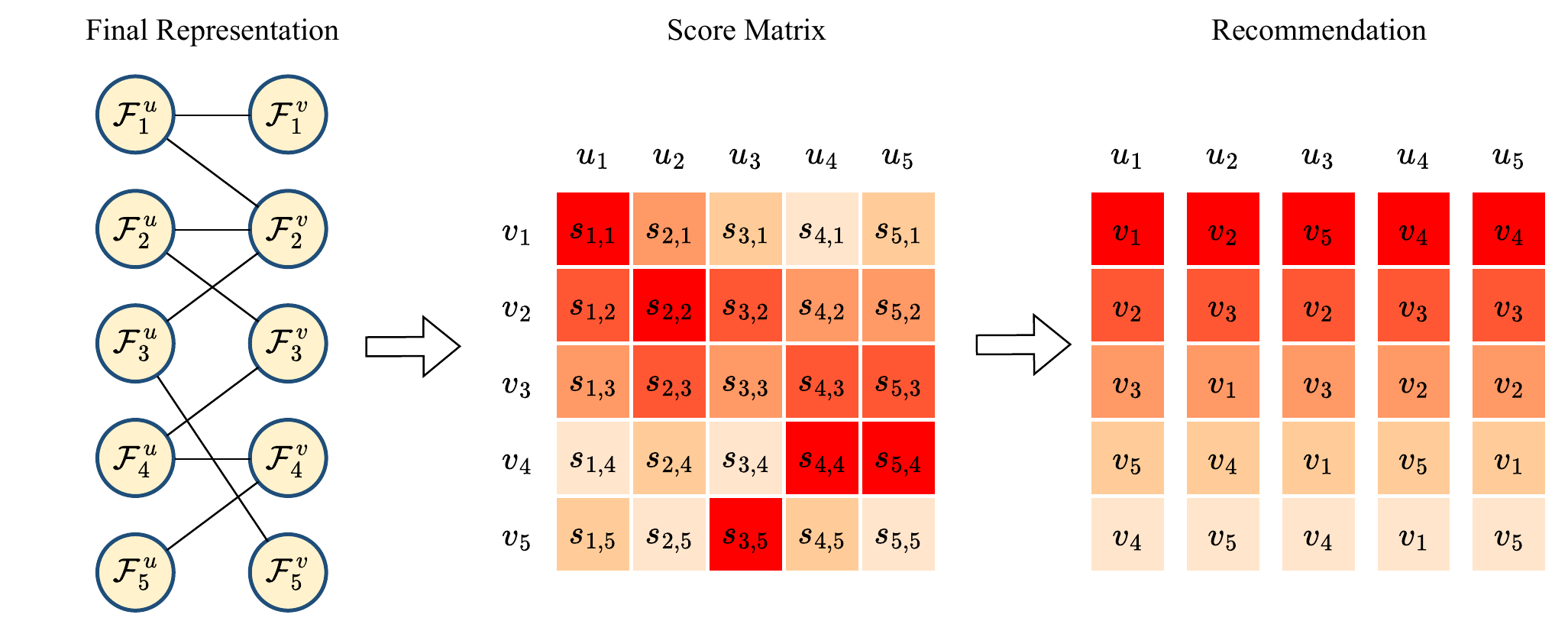}
  \caption{The final step of the proposed pipelines consists in providing personalized recommendations to users. This involves computing a score $s_{i,j}$ for each user $i$ and item $j$ by performing matrix multiplication between the user embedding $\mathcal{F}^u$ and the item embedding $\mathcal{F}^v$: $(\mathcal{F}^u)^T\mathcal{F}^v$. Finally, scores are ranked to present each user with the most relevant items.}
  \label{fig:model_rec}
\end{figure}

%Before training, we associate two nodes $u,v$ with their respective users. %Then, $\mathcal{F}(U_v)$ (resp. $\mathcal{F}(U_u)$) is the vector space describing all the possible facets of $V$ (resp. $U$). We remark that these two stalks represent the respective users, and instead of a vector, we exploit the richer structure of a vector space. 
%Each node is split into two partitions. One containing positive samples, i.e., pairs of nodes where interactions were positive (meaning, user positively rating an item), and one containing negative samples, which define instances not rated by the user.
%$e$ is the edge representing the shared characteristics (for instance, all items commonly liked or interacted with) between the two users. The item's features are modeled into the vector space $\mathcal{X}(U_e)$. 
%At each iteration of the training loop, the model updates these features for all the users.

The last sheaf layer outputs the restriction maps $\mathcal{F}_{u \unlhd e}$ and $\mathcal{F}_{v \unlhd e}$, which represent the final user and item representations, respectively. These maps, denoted without the $\unlhd e$ for simplicity, are matrices of shapes $d_e \times d_u$ and $d_e \times d_v$.

To rank items for each user $i$, we need a score $s_{i,j}$ that specifies the relevance of item $j$ for user $i$. This can be achieved by multiplying the final representation user $i$, $\mathcal{F}^u_i$, with the final representation of item $j$, $\mathcal{F}^u_j$. This operation is repeated for each user $i$ and item $j$ in the graph, and can be expressed as a matrix multiplication between the user embeddings $\mathcal{F}^u$ and the item embeddings $\mathcal{F}^v$: $(\mathcal{F}^u)^T\mathcal{F}^v$. The final result is a score matrix $S$ of the form $d_u \times d_v$, containing the scores given to each user-item pair $(i,j)$.

These scores are then ranked to present each user $i$ with a tailored list of the most relevant items.

This computation process is illustrated in Figure \ref{fig:model_rec}.

\subsection{Training}
%%Pure descrizione della loss e di come viene adattata al metodo
\begin{figure}[t]
  \centering
  \includegraphics[width=\textwidth]{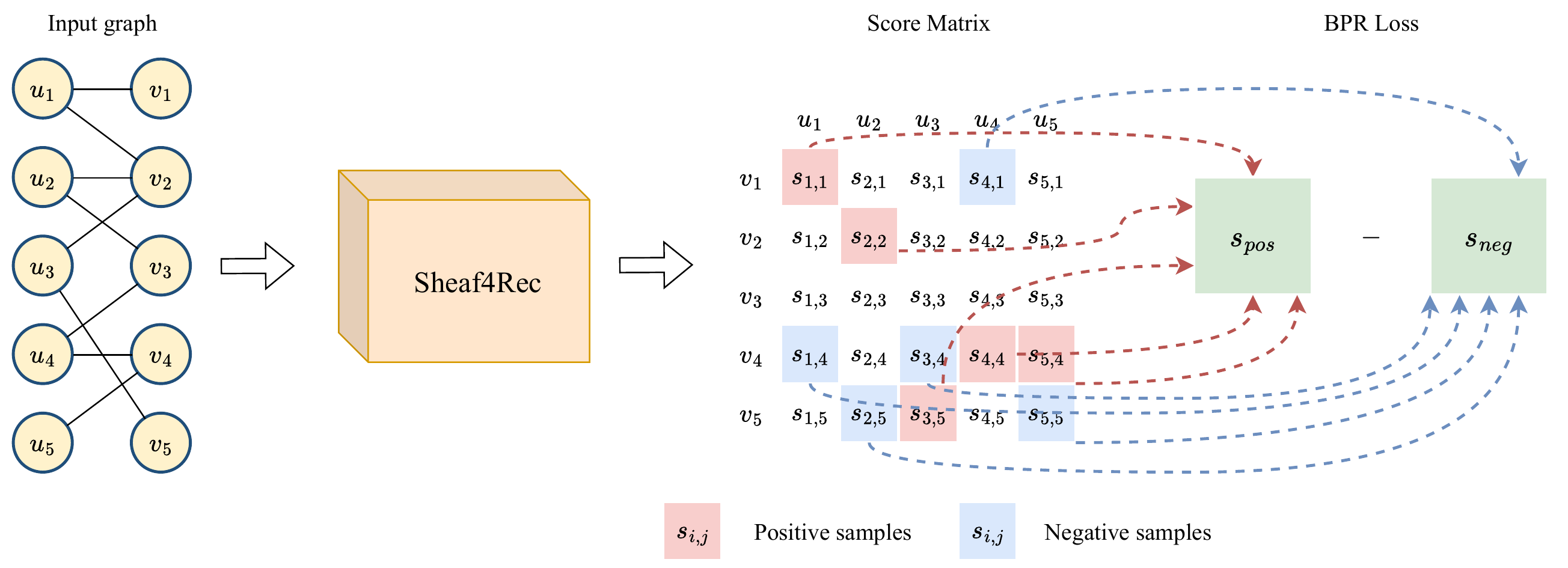}
  \caption{During training, \ModelName takes as input the user-item graph, and generates a score matrix $S$ that encapsulates the computed relevance of each item $j$ for each user $i$. To compute the loss function, it is necessary to extract from this matrix the scores corresponding to both positive items ((those with which the user has already interacted) and negative items (those with which the user has not yet interacted). These scores are then used to compute the two terms of the BPR Loss, which is minimized during each training iteration.}
  \label{fig:model_training}
\end{figure}

The Bayesian Personalized Ranking (BPR) loss is a natural choice for the loss function in recommender systems \cite{https://doi.org/10.48550/arxiv.1205.2618}. To understand the essence of BPR, it's important to define the concepts of positive and negative edges. Positive edges correspond to user actions, such as purchasing or selecting an item, and are therefore present in the graph. In contrast, negative interactions are not observed, resulting in the abscence of the edges in the graph.

Consequently, we define the set of positive scores $S_{pos} = \{s_{i,j}, s.t. (i,j) \in E, i \in U, j \in V\}$, which includes all scores corresponding to edges present in the graph. Conversely, the set of negative scores $S_{neg} = \{s_{i,j}, s.t. (i,j) \notin E, i \in U, j \in V\}$ comprises all scores corresponding to edges not present in $E$.

The BPR loss is then calculated by summing all the positive scores and negative scores, and subtracting the latter from the former:
\begin{equation}
\label{bpr_loss}
    BPR(S) = -ln\left(\sigma(\sum_{s_{i,j} \in S_{pos}}s_{i,j}-\sum_{s_{i,j} \in S_{neg}}s_{i,j})\right) = -ln\left(\sigma(s_{pos}-s_{neg}\right)
\end{equation}
where $S$ is the score matrix, $s_{pos}$ and $s_{neg}$ represent the two summation terms, respectively.

Due to computational complexity, we use mini-batch sampling during training to estimate the BPR loss. For each mini-batch, we sample a subset of users, and for each user, we randomly sample one positive item and one negative item. This training strategy emphasizes the probability of a user favoring a positively observed item over a randomly chosen negative counterpart.

The training process is exemplified in Figure \ref{fig:model_training}.

\section{Experiments}
\label{experiments}
Our experiments on the performance of 
\ModelName on collaborative filtering aim to answer the
following Research Questions: 
\begin{itemize}
    \item \textbf{RQ1}: How does \ModelName perform against state-of-the-art
    baselines?
    \item \textbf{RQ2}: What is the impact of hyperparameters on recommendation performance?
    \item \textbf{RQ3}: Does the categorical structure of sheaf architecture introduce more expressive power?
\end{itemize}

\subsection{Datasets description}

\begin{table}[ht]
  \begin{tabular}{cccc}
    \toprule
     & Facebook Books & Yahoo! Movies & MovieLens 1M \\ 
    \midrule
    Users & 1398 & 4000 & 6040 \\
    Items & 2933 & 2626 & 3900\\
    Ratings & 1878 & 69846 & 1000000\\
    Density [\%] & 0.025 & 0.664 &  4.24 \\
    \bottomrule
  \end{tabular}
  \caption{Number of users, number of items, number of ratings and density (percentage) for each dataset.}
\label{dataset_info}
\end{table}

We use three datasets: MovieLens 1M\footnote{MovieLens 1M Dataset: \url{https://grouplens.org/datasets/movielens/1m/}}, Yahoo! Movies\footnote{Yahoo! Movies Dataset: \url{https://
webscope.sandbox.yahoo.com/catalog.php?datatype=r}} and Facebook Books\footnote{Facebook Books Dataset: \url{https://2015.eswc-conferences.org/important-dates/call-RecSys.html}}. We decided to use these datasets because they have been widely used in previous research \cite{10.1145/3604915.3608804,10.1145/3588901} and offer different characteristics suitable for evaluating our model. Their main characteristics are presented in Table \ref{dataset_info}.

Here's an overview of each dataset:
\begin{itemize}
    \item \textbf{Facebook Books:}
    The Facebook Books dataset provides a relatively sparse collection of explicit feedback related to books. It contains 1,878 comments from 1,398 people about 2,933 products (books).
    
    \item \textbf{Yahoo! Movies:}
    The Yahoo! Movies dataset contains 69,846 explicit movie ratings on a scale of 1 to 5, gathered from the Yahoo! Movies website as of November 2003. It contains information about 2,626 items and includes 4,000 user reviews. %Similarly to the Facebook Book dataset, explicit feedback is converted to implicit using a thresholding techinique: review scores of $3$ or higher are classified as positive, otherwise, they are marked as negative.
    
    \item \textbf{MovieLens 1M:}
    The MovieLens 1M dataset comprises around 1 million ratings provided by 6,040 users for approximately 3,900 movies. Each user has rated at least 20 movies on a 5-star scale with half-star increments. Demographic information such as age, gender, and occupation is also available for each user.
\end{itemize}

\subsection{Baselines}
We compare our model to several baselines, including both established state-of-the-art models and recently proposed neural architectures. Here's a brief summary of the baselines used in our experiments.

\begin{itemize}
    \item \textbf{NGCF:}
    Neural Graph Collaborative Filtering (NGCF) \cite{Wang2019-gl} integrates user-item interactions into the embedding process, incorporating collaborative signals directly into the embedding computation.
    \item \textbf{LightGCN:}
    LightGCN \cite{He2020-pe} is a variant of the Graph Convolutional Neural network (GCN) \cite{https://doi.org/10.48550/arxiv.1609.02907}, which includes neighborhood aggregation for collaborative filtering.
    \item \textbf{GAT:}
    We adapted the popular Graph Attention Network (GAT) architecture \cite{https://doi.org/10.48550/arxiv.1710.10903}, which exploits attention mechanisms to capture relational dependencies, to serve as a collaborative filtering method on graph data.
    \item \textbf{UltraGCN:}
    UltraGCN \cite{mao2023ultragcn} provides a mathematical proxy to the GCN \cite{https://doi.org/10.48550/arxiv.1609.02907} layer to mitigate the negative impact of over-smoothing in recommendation tasks.
    \item \textbf{KGTORe:}
    Knowledge-Graph and Tree-Oriented Recommendation (KGTORe) \cite{10.1145/3604915.3608804} architecture leverages a knowledge graph to learn latent representations for semantic features, interpreting user decisions as a personal distillation of the item feature representations.
\end{itemize}

\subsection{Metrics}
In our evaluation, we employ three frequently adopted metrics from previous work \cite{10.1145/3604915.3608804, mao2023ultragcn} to assess the performance of recommendation systems::
\begin{itemize}
    \item \textbf{F1-Score@K:} This metric calculates the harmonic mean of precision and recall at rank K. Precision measures the proportion of recommended items relevant to the user, while Recall quantifies the proportion of relevant items successfully recommended. The harmonic mean balances Precision or Recall, ensuring neither dominates the score disproportionately.
    \item \textbf{NDCG@K:} Normalized Discounted Cumulative Gain (NDCG) measures the effectiveness of a ranking system by taking into account the position of K relevant items in the ranked list. It assigns higher scores to items ranked higher in the list, reflecting their perceived importance.
    \item \textbf{MRR@K:} The Mean Reciprocal Rank (MRR) calculates the average reciprocal of the rank at which the first relevant recommendation appears.
\end{itemize}

In addition, we report the time required to generate $100$ recommendations as an evaluation criterion.

\subsection{Reproducibility}

\begin{table}[ht]
  \begin{tabular}{ccc}
    \toprule
    & Hyperparameter & Value\\ 
    \midrule
    $l$ & Latent dimensions & $64$ \\
    $N$ & Number of layers & $5$ \\
    & Learning rate & $10^{-3}$\\
    & Number of epochs & 100 \\
    & Batch size & 1024 \\
    & Weight decay & $10^{-4}$ \\
    \bottomrule
  \end{tabular}
  \caption{Hyperparameter values for our model architecture.}
\label{table:hparams}
\end{table}

We ensure the reproducibility of our experiments by providing access to our codebase, datasets and detailed procedures on our Github Repository\footnote{\url{https://github.com/antoniopurificato/Sheaf4Rec}}.

To partition the datasets, we randomly select 80\% of each user's past interactions for training, reserving 10\% for testing. We allocate the remaining 10\% of interactions for validation in order to effectively tune the hyper-parameters \cite{He2020-pe, Wang_2018}.

We use Bayesian hyperparameter optimisation, exploiting a search space defined by optimal parameters identified in previous studies. Six distinct hyperparameter configurations are explored. The optimal hyperparameters used to train \ModelName are outlined in Table \ref{table:hparams}.

Our experiments are performed on a workstation equipped with an Intel Core i9-10940X (14-core CPU running at 3.3GHz), 256GB of RAM, and a single Nvidia RTX A6000 with 48GB of VRAM.
\section{Results}
\label{results}

\newcommand{\mydag}{\textsuperscript{\textdagger}}
To answer the research questions and evaluate the effectiveness of \ModelName, we compare its performance against recent baselines. Subsequently, we perform hyperparameter and ablation tests in order to evaluate our components.

\subsection{Performance Comparisons with Baselines (RQ1)}

\begin{table*}[!ht]
    \centering
    \resizebox{\textwidth}{!}{
    \begin{tabular}{c|cccc|cccc|cccc}
        \toprule
         &    \multicolumn{4}{c|}{Facebook}     &    \multicolumn{4}{c|}{Yahoo}    &    \multicolumn{4}{c}{MovieLens 1M}\\
         \midrule
         & F@10    & N@10 & F@20 & N@20 & F@10 & N@10 & F@20 & N@20 & F@10        & N@10 & F@20 & N@20 \\
         \midrule
GAT      & 0.019\mydag    & 0.036\mydag   & 0.018\mydag & 0.051\mydag   & 0.069\mydag & 0.119\mydag   & 0.055\mydag & 0.142\mydag   & 0.109\mydag        & 0.143\mydag   & 0.122\mydag & 0.156\mydag   \\
NGCF     & 0.027\mydag    & \underline{0.062}   & \underline{0.022}\mydag & \underline{0.078}\mydag   & \underline{0.073}\mydag & 0.129\mydag   & \underline{0.057}\mydag & 0.155   & \underline{0.129}\mydag        & \underline{0.169}\mydag   & \underline{0.141}\mydag & \underline{0.182}\mydag   \\
LightGCN & 0.002\mydag    & 0.001\mydag   & 0.020\mydag & 0.003\mydag   & 0.032\mydag & 0.120\mydag   & 0.035\mydag & 0.070\mydag   & 0.003\mydag  & 0.004\mydag   & 0.004\mydag & 0.005\mydag   \\
KGTORe   & 0.022\mydag    & 0.035\mydag   & 0.019\mydag & 0.044\mydag   & 0.071\mydag & \underline{0.137}\mydag   & 0.052\mydag & \underline{0.159}\mydag   & 0.076\mydag        & 0.125\mydag   & 0.089\mydag & 0.130\mydag   \\
UltraGCN & \underline{0.027}\mydag    & 0.061\mydag   & 0.019\mydag & 0.066\mydag   & 0.057\mydag & 0.101\mydag   & 0.048\mydag & 0.117\mydag   & 0.083\mydag        & 0.147  & 0.104\mydag & 0.155\mydag   \\
\ModelName (ours)  & \textbf{0.029}    & \textbf{0.066}   & \textbf{0.024} & \textbf{0.081}   & \textbf{0.076} & \textbf{0.147}   & \textbf{0.062} & \textbf{0.162}   & \textbf{0.140}        & \textbf{0.182}   & \textbf{0.151} & \textbf{0.197      }  \\
\midrule
Improvement (\%)&  7.41 & 11.29 & 9.09 & 3.85 & 4.11 & 7.30 &  8.77 & 1.82 & 8.53 & 7.69 & 7.09 & 8.24\\
\bottomrule
\end{tabular}
}
\caption{Performance comparison of the six models on the three tested datasets, measured in terms of NDCG@K (N) and F1@K (F),with $K=\{10,20\}$. Statistically significant differences betweeen \ModelName and the baselines, determined by paired Wilcoxon tests ($p < 0.01$) with Bonferroni correction, are marked with \textdagger. Our approach consistently outperforms all baselines in terms of both metrics across all datasets.}
%The highest relative improvement is obtained on the Facebook Books dataset in terms of NDCG@10 with an 11.29\% improvement with respect to NGCF, the second-best model.}
\label{main_results_table}
\end{table*}

Table \ref{main_results_table} presents a comprehensive overview of the performance comparison between \ModelName and the competing baselines, showing their performance on the test set. We highlight the best result in \textbf{bold} and second best result with \underline{underline}.

Statistical validation of the results was performed using paired Wilxocon tests with a significance level of $0.01$ and Bonferroni correction \cite{weisstein2004bonferroni}. Significant improvements are marked with \textdagger.

It is worth noting that models based on graph architectures, i.e. LightGCN, NGCF, and UltraGCN, regularly achieve competing performance on all three datasets. The same holds for models based on knowledge graphs, such as KGTORe, which incorporate additional side information, like user gender or age.
In terms of NDCG, \ModelName surpasses all baselines. In particular, \ModelName achieves an impressive $11.29\%$ improvement in NDCG@10 over NGCF on the Facebook Books dataset. Similarly, \ModelName shows an $8.29\%$ improvement in NDCG@20 on the MovieLens 1M dataset compared to NGCF, which is the second-best performing model.

\begin{table*}[!ht]
    \centering
    \resizebox{\textwidth}{!}{
    \begin{tabular}{c|cccc|cccc|cccc}
        \toprule
         &    \multicolumn{4}{c|}{Facebook}     &    \multicolumn{4}{c|}{Yahoo}    &    \multicolumn{4}{c}{MovieLens 1M}\\
         \midrule
         & P@10 & R@10 & P@20 & R@20 & P@10 & R@10 & P@20 & R@20 & P@10 & R@10 & P@20 & R@20  \\
         \midrule
GAT     &  0.015\mydag &  0.096\mydag &  \textbf{0.013}\mydag &  \textbf{0.155}\mydag &  0.036\mydag &  0.155\mydag &  0.028\mydag &  0.225\mydag &  0.111\mydag &  0.103\mydag &  0.087\mydag &  0.153\mydag\\
NGCF     &  0.016 &  0.097\mydag &  0.011\mydag &  0.143\mydag &  0.044\mydag &  \underline{0.191}\mydag &  \underline{0.033} &  \underline{0.274}\mydag &  0.129\mydag &  \underline{0.123}\mydag &  0.106\mydag &  \underline{0.192}\mydag\\
LightGCN &  0.001\mydag &  0.002\mydag &  0.001\mydag &  0.004\mydag &  0.001\mydag &  0.003\mydag &  0.001\mydag &  0.014\mydag &  0.002\mydag &  0.001\mydag &  0.002\mydag &  0.002\mydag\\
KGTORe   &  \textbf{0.021}\mydag &  \underline{0.100}\mydag &  0.011\mydag &  0.082\mydag &  0.044\mydag &   0.189\mydag  & 0.031\mydag & 0.221\mydag & \underline{0.153}\mydag &  0.063\mydag &  \textbf{0.122} &  0.101\mydag \\
UltraGCN &  \underline{0.020}\mydag &  0.060\mydag &  0.011\mydag &  0.910\mydag &  \underline{0.045}\mydag &  0.143\mydag &  0.019\mydag &  0.079\mydag &  \textbf{0.170}\mydag &  0.072\mydag &  0.144\mydag &  0.120\mydag\\
\ModelName (ours)  & 0.017 &  \textbf{0.104} &  \underline{0.012} &  \underline{0.147} &  \textbf{0.047} &  \textbf{0.202} &  \textbf{0.034} &  \textbf{0.286} &  0.141 &  \textbf{0.138} &  \underline{0.116} &  \textbf{0.214}\\
\bottomrule
\end{tabular}}
\caption{Performance comparison of the six models on the three tested datasets, measured in terms of Precision@K (P) and Recall@K (R),with $K=\{10,20\}$. Statistically significant differences betweeen \ModelName and the baselines, determined by paired Wilcoxon tests ($p < 0.01$) with Bonferroni correction, are marked with \textdagger. \ModelName is able to obtain a good trade-off between precision and recall, while competitors, like UltraGCN, have high precision and low recall, indicating some problems in their recommendation.}
\label{table:precision_and_recall}
\end{table*}

In terms of F1, our model consistently outperforms all the competing baselines across all the evaluated datasets. As can be seen from Table \ref{table:precision_and_recall}, the main strength of our approach is its ability to strike a good trade-off between precision and recall. Some approaches, like LightGCN, have high precision and low recall, suggesting that the system is cautious in suggesting items and it might not cover the entire breadth of relevant items available. Other approaches, such as KGTORe, show the opposite trend, emphasizing recall at the expense of precision, indicating a system that manages to include a lot of potentially relevant items in its recommendations, but also many items that do not align well with the user's preferences or needs.

We attribute the superiority of our model to several factors, including:
\begin{itemize}
    \item Improved expressiveness: Our model, employing cellular sheaves, offers a higher degree of expressiveness, which allows it to capture complex user-item relationships more effectively.
    \item Better alignment with the recommendation domain: The inherent characteristics of a sheaf-based approach are likely to be better aligned with the nuances of recommendation tasks, resulting in more accurate and relevant recommendations.
\end{itemize}

\begin{table*}[!ht]
    \centering
    \begin{tabular}{c|cc|cc|cc}
        \toprule
         &    \multicolumn{2}{c|}{Facebook}     &    \multicolumn{2}{c|}{Yahoo}    &    \multicolumn{2}{c}{MovieLens 1M}\\
         \midrule
         & MRR@10    & MRR@20 & MRR@10 & MRR@20  & MRR@10 & MRR@20 \\
         \midrule
GAT      & 0.037\mydag & 0.032\mydag & 0.066\mydag    & 0.071\mydag   & 0.043\mydag & 0.038\mydag    \\
NGCF     & 0.037\mydag & 0.049 & 0.072\mydag   & 0.077   & \underline{0.060}\mydag & \underline{0.054}\mydag   \\
LightGCN & 0.005\mydag    & 0.006\mydag   & 0.006\mydag & 0.007\mydag   & 0.004\mydag & 0.005\mydag \\
KGTORe   & 0.036\mydag & \textbf{0.054}\mydag & \underline{0.081}\mydag    & \underline{0.084}\mydag   & 0.055\mydag & 0.044\mydag  \\
UltraGCN & \underline{0.041}\mydag    & 0.045\mydag   & 0.070\mydag & 0.083\mydag   & 0.057\mydag & 0.047\mydag\\
\ModelName (ours) & \textbf{0.044}    & \underline{0.051}   & \textbf{0.085} & \textbf{0.090}  & \textbf{0.067} & \textbf{0.059}\\
\midrule
Improvement (\%)&  7.32 & -5.88 & 4.94 & 7.14 & 11.67 & 9.26\\
\bottomrule
\end{tabular}
\caption{MRR@10 and MRR@20 of our approach compared with the other approaches on the three datasets. Our solution outperforms all the baselines on all the metrics, except for KGTORe on Facebook Books. The results which are statistically significant based on paired Wilcoxon tests ($p < 0.01$) with Bonferroni correction are denoted with \textdagger. On the Facebook Books dataset, \ModelName has a 7.32 \% improvement with respect to UltraGCN, the second-best model.}
\label{mrr_results_table}
\end{table*}

In terms of MRR, our solution shows notable improvements over all baseline models, with the exception of KGTORe on the Facebook dataset. The higher MRR values indicate our model's ability to prioritize relevant recommendations early in the ranking, potentially boosting user satisfaction and overall experience with the recommendation system.
This positive outcome can be attributed to our model's ability to learn an adjacency matrix based on node features. By establishing a robust correlation between the underlying graph structure and node attributes, our model sorts items effectively, enhancing their positioning in the recommendation list. %The Reciprocal Rank, computed using the first K relevant items in the recommendation list, benefits from the relationship our model establishes between the graph structure and node attributes, ensuring more accurate item sorting.

\subsection{Ablation Studies (RQ2)}

\subsubsection{Values of K}
\begin{figure}[!ht]
  \centering
  \includegraphics[width=\linewidth]{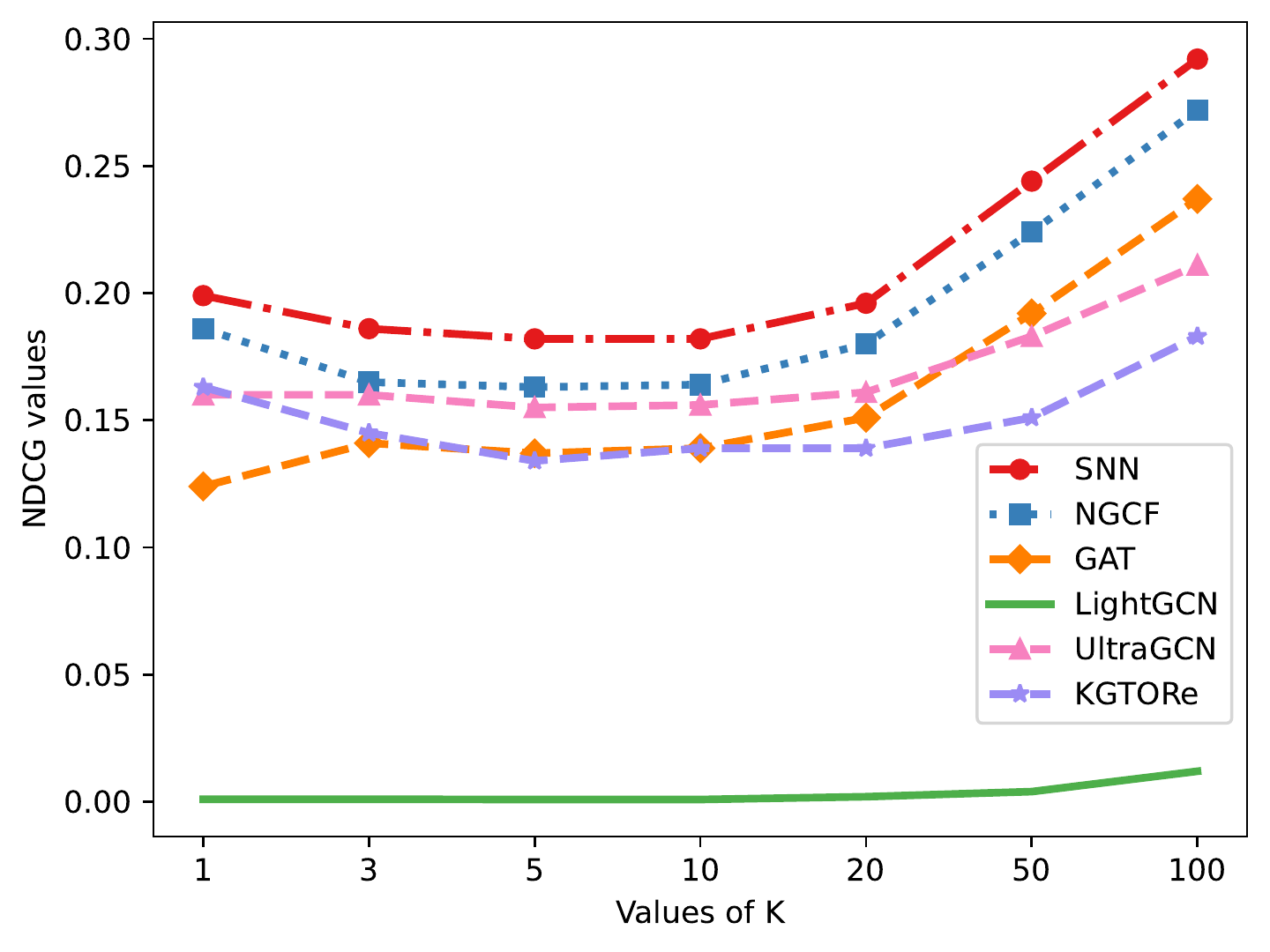}
  \caption{NDCG@K across different values of K for the six models tested on the ML-1M dataset. \ModelName consistently outperforms other baselines, maintaining its superiority as the parameter K increases.}
  \label{ndcg_for_K_values}
\end{figure}

In Figure \ref{ndcg_for_K_values}, we analyse the performance in terms of NDCG of \ModelName alongside other state-of-the-art systems as we vary the parameter $K$, i.e. the number of recommendations generated. It's evident that \ModelName consistently surpasses the baselines, as its NDCG@K remains superior to all other methods across all values of $K$. 

In particular, outperforming baselines at higher values of $K$, showcases our system's ability to scale effectively. As the number of items to recommend increases, our system maintains or even improves its recommendation quality, which is crucial in scenarios where extensive top-K recommendations are required.

\subsubsection{Recommendation Time}
Table \ref{recommendation_time} provides a detailed comparison of the recommendation time between \ModelName and other state-of-the-art methods. To ensure fairness in our evaluation, we maintained consistency by using the same number of layers for each network (2 and 5 layers) and keeping the hyperparameters constant. The computation time, measured in seconds, reveals the remarkable efficiency of \ModelName, as it consistently outperforms all other methods, showcasing the shortest computational time. 
Despite its inherently complex topological structure, our method remains stable and consistently achieves top-tier performance. To further highlight the stability of our solution, we present the standard deviation computed over 10 attempts. Notably, \ModelName's standard deviations are persistently lower than those of the other methods, demonstrating great consistency and stability.
Strikingly, LightGCN shows significant instability, as evidenced by the higher standard deviation values in Table \ref{recommendation_time}.

\begin{table}[ht]
  \begin{tabular}{cccc}
    \toprule
    Model & Facebook Rec Time [s] & Yahoo Rec Time [s] & MovieLens 1M Rec Time [s]\\ 
    \midrule
    GAT & 0.594 $\pm$ 0.090 & 1.344 $\pm$ 0.077 & 3.796 $\pm$ 0.068\\
    NGCF & 0.607 $\pm$ 0.057 & 1.329 $\pm$ 0.073 & 3.821 $\pm$ 0.060\\
    LightGCN & 0.804 $\pm$ 0.349 & 1.434 $\pm$ 0.377 &  \textbf{3.633} $\pm$ 0.297\\
    KGTORe & 0.763 $\pm$ 0.110 & 1.458 $\pm$ 0.066 & 4.001 $\pm$ 0.035 \\
    UltraGCN & 1.278 $\pm$ 0.044& 1.388 $\pm$ 0.111 & 3.823 $\pm$ 0.076\\
    \ModelName (ours) & \textbf{0.586} $\pm$ 0.045  & \textbf{1.309} $\pm$ 0.045 & 3.770 $\pm$ 0.051\\
    \bottomrule
  \end{tabular}
  \caption{Time (in seconds) required to provide the top 100 recommendations to a user compared to all the other baselines.
  %$\downarrow$ is better.
    Our solution outperforms all approaches on two of the three datasets. On the third dataset, ML-1M, LightGCN surpasses \ModelName, but its standard deviation is remarkably high.}
  %When LightGCN reaches the peak of the standard deviation, our \ModelName performs better.}
\label{recommendation_time}
\end{table}

\subsubsection{Impact of the BPR Loss}
We investigate the effectiveness of different popularly used loss functions in training \ModelName, comparing the Bayesian Personalised Ranking (BPR) loss against the Root Mean Squared Error (RMSE) and the Binary Cross-Entropy (BCE) losses. We compute F1-Score@10 on the MovieLens 1M dataset, as summarised in  Table \ref{comp_time}.

Our results show that \ModelName trained with BPR loss achieves better performance than trained using RMSE or BCE. To ensure a complete analysis, we varied the number of layers in each architecture to mitigate any confounding effect of network depth on performance.

This investigation emphasizes the intrinsic relationship between the sheaf structure and the BPR loss, as discussed in Section \ref{method}, underscoring the importance of matching the model architecture with an appropriate loss function, ultimately optimising performance and effectiveness in real-world applications.

Additionally, the last column of Table \ref{comp_time} highlights that BPR has a considerable computational edge over RMSE and BCE in terms of the training time.

\begin{table}[ht]
  \begin{tabular}{c|cccccc}
    \toprule
    Loss & \#Layers & F1@10 & Time [min]\\ 
    \midrule
    \multirow{2}{*}{RMSE} & 2 & 0.112 & 283\\
     & 5 & 0.118 & 371\\
    \midrule
    \multirow{2}{*}{BCE} & 2 & 0.087 & 198\\
     & 5 & 0.101 & 262\\
    \midrule
    \multirow{2}{*}{BPR} & 2 & 0.123 & \textbf{163}\\ 
     & 5 & \textbf{0.140} & 210\\
    \bottomrule
  \end{tabular}
  \caption{Performance in terms of F1-Score@10 and training time (in minutes) of \ModelName on ML-1M when using different loss functions and numbers of layers. BPR Loss obtains both the highest F1@10 and the lowest training time, for both number of layers.
  %$\downarrow$ is better.
  }
\label{comp_time}
\end{table}

\subsubsection{Latent dimensions analysis}
Our architecture, as discussed in Section \ref{method}, allows stacking multiple layers to propagate new user and item features, facilitating the creation of distinct clusters of users with similar preferences. The update of each latent feature vector as a function of its local neighbourhood is determined by a parameter called \textit{latent dimension} $l$. This parameter represents the dimensionality of the latent space within each sheaf layer, thereby affecting the size of the learned representations and transformations within the network, ultimately influencing the expressiveness and capacity of the model.

\begin{figure}[ht]
  \centering
  \includegraphics[width=\textwidth]{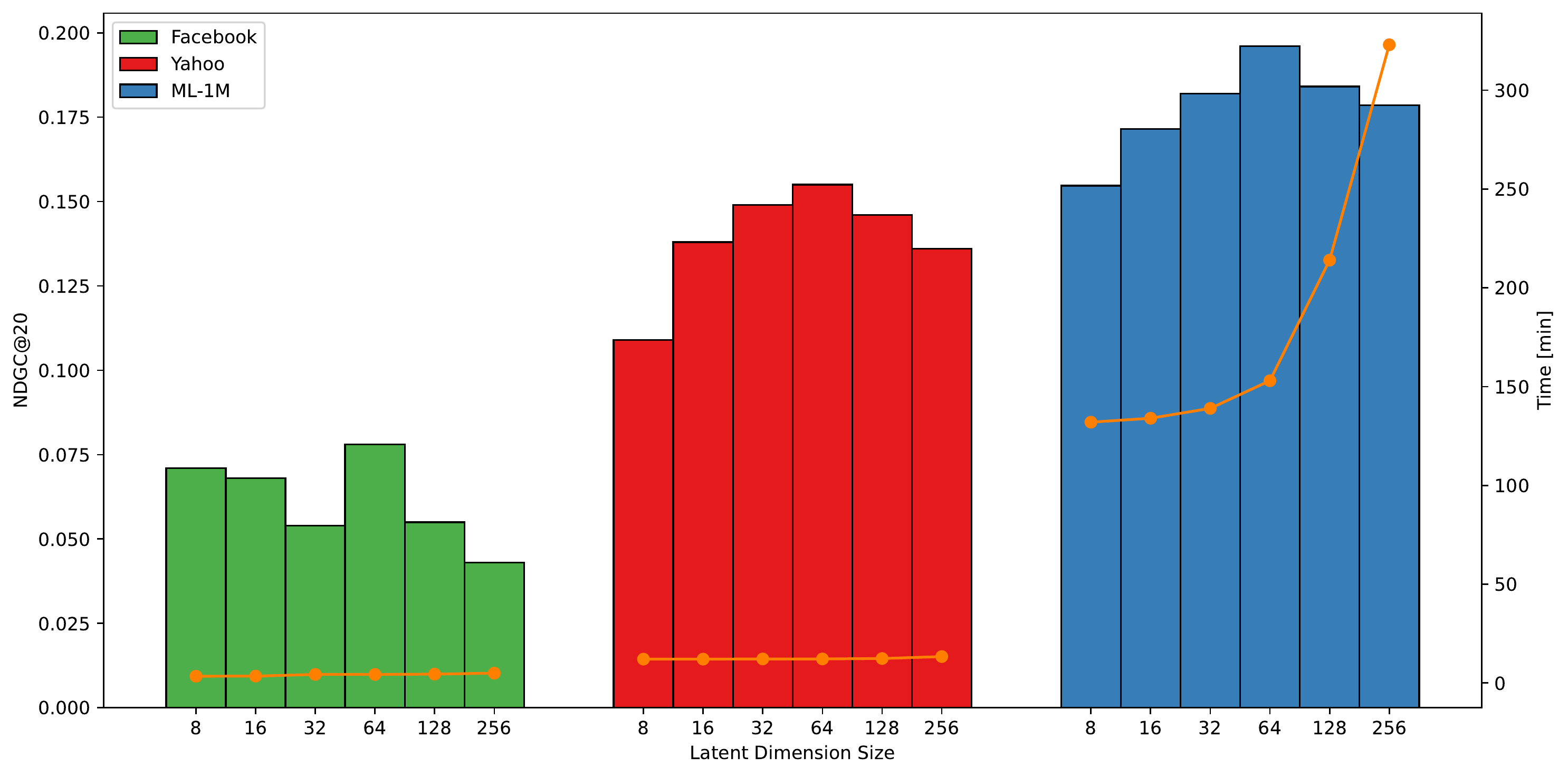}
  \caption{NDCG@20 for \ModelName across three datasets while varying the latent dimension size. The computational time is shown on a secondary y-axis.
  Optimal performance is achieved when using a latent dimension of 64. As expected, as the latent dimension increases, so does the computational time.}
  \label{latent_dim_variation_plot}
\end{figure}

We tested 5 different values for $l$, as depicted in Figure \ref{latent_dim_variation_plot}, and found that a latent dimension of 64 yields the best results across all the datasets. From the plot, we can draw several insights:
\begin{itemize}
    \item \textbf{Optimal latent dimension $l$:} Increasing the latent dimension beyond a certain threshold negatively affects performance, indicating the importance of finding a suitable trade-off based on the specific task.
    \item \textbf{Computational considerations:} Higher latent dimension corresponds to increased computational time, especially evident for larger datasets such as MovieLens 1M. This phenomenon is due to the larger latent space, which in turn increasese the size and complexity of the model, leading to longer training and inference times.
\end{itemize}

In essence, the choice of $l$ serves as a critical design parameter, directly affecting both the performance and computational efficiency of the model. 

\subsection{The advantages of the categorical structure (RQ3)}
\subsubsection{Number of Layers}
\label{sec:layers}
One of the distinctive advantages of \ModelName lies in its capability to mitigate over-smoothing by incorporating numerous sheaf layers \cite{https://doi.org/10.48550/arxiv.2202.04579}. In contrast, several GNN architectures have shown a decrease in performance as the layer count increases \cite{https://doi.org/10.48550/arxiv.1905.1094}.

\begin{figure}[t]
  \centering
  \includegraphics[width=\textwidth]{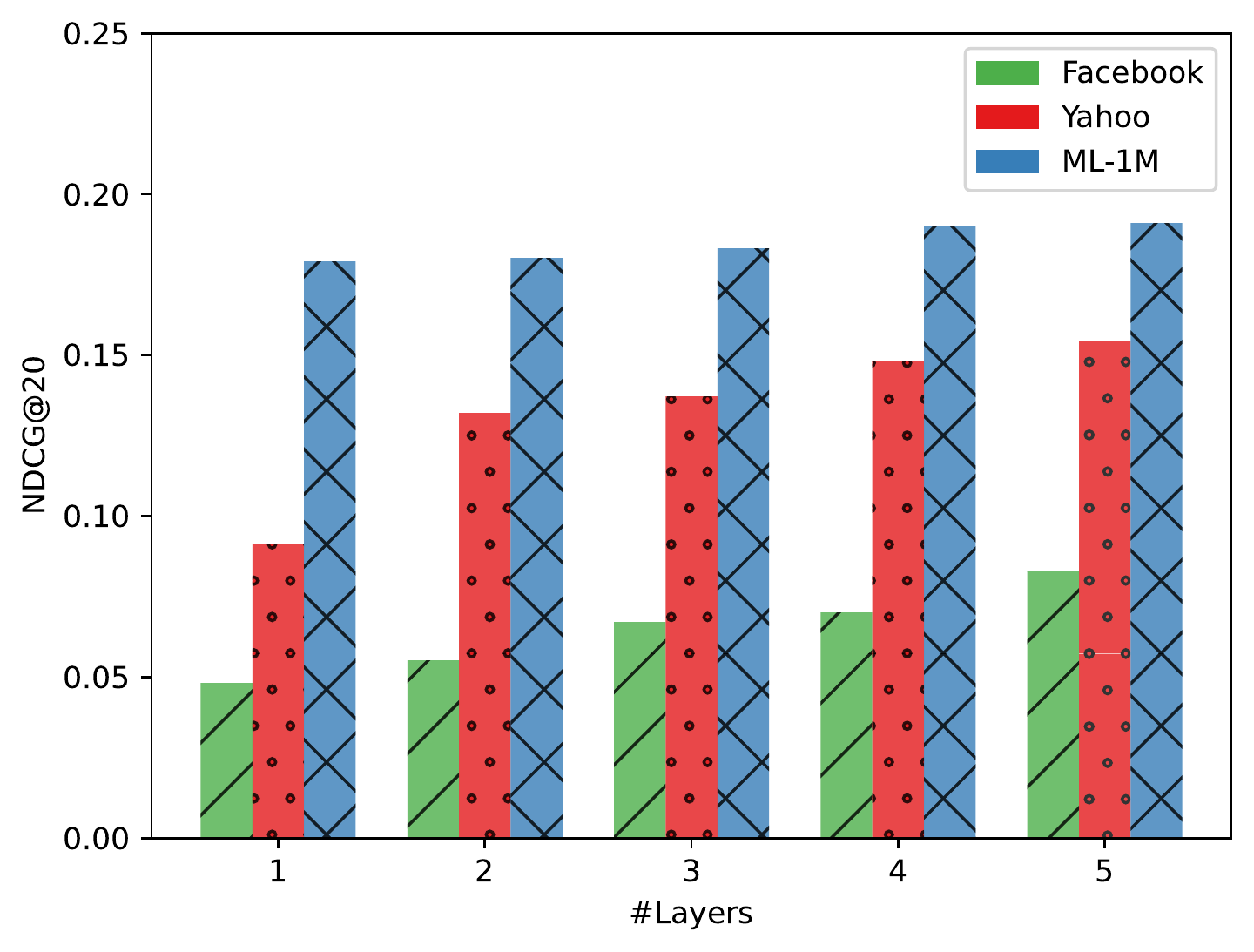}
  \caption{NDCG@20 values of \ModelName across three datasets depending on the number of layers. For all the dataset the best results are obtained using 5 layers, highlighting that our model is robust to over-smoothing problems commonly encountered in GNNs.}
  \label{5_layers}
\end{figure}

To assess the impact of the number of layers on performance, we evaluated the NDCG@20 across varying layer counts, ranging from $N=2$ to $N=5$. As depicted in Figure \ref{5_layers}, optimal performance tends to be achieved when utilizing $N=5$ layers. This empirically finding underscores the effectiveness of sheaf diffusion processes in mitigating oversmoothing, as further elucidated in \cite{https://doi.org/10.48550/arxiv.2202.04579}.

\subsubsection{Expressive power of the proposed formalism}

\begin{figure}[!ht]
  \centering
  \includegraphics[width=\linewidth]{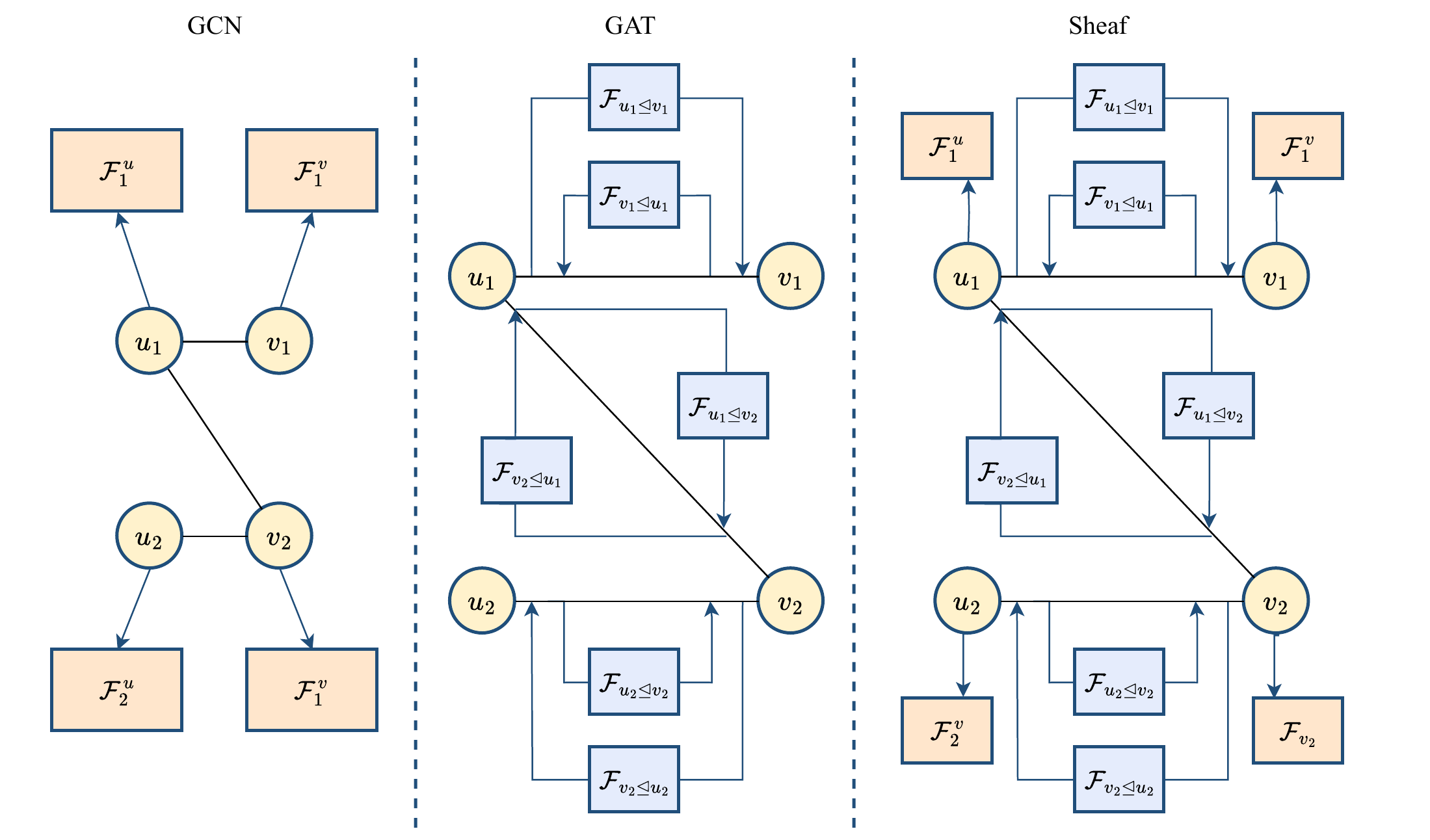}
  \caption{GCN, GAT and Sheaf architectures, differentiated by their node and edge stalks configurations. In the GCN architecture (left), the edge size is set to $1$, while the size of node stalks is $N$. By setting to $1$ the size of node stalks and to $N$ the size of edge stalks, we characterise the GAT architecture (center). The Sheaf architecture (right) is obtained when both node and edge stalks sizes are equal to $N$}
  \label{fir:gat-gcn-sheaf}
\end{figure}

The objective of this experiment is to empirically showcase the expressive capacity of the sheaf architecture. By changing the size of node and edge stalks, we investigate three different configurations:

\begin{itemize}
    \item \textbf{GAT equivalent:} $\dim{(\mathcal{F}(v))}=1$ and $\dim{(\mathcal{F}(e))}=N$. This configuration mirrors the GAT architecture \cite{https://doi.org/10.48550/arxiv.1710.10903}. In GAT, the vector space associated to nodes is one-dimensional, allowing each node to attend every other node, ignoring structural information. Edge stalks have size $N$, allowing to compute the self-attention for each pair of nodes, a feature unattainable with edge stalks of size $1$, which limits self-attention computation only to neighbouring nodes.
    \item \textbf{GCN equivalent:} $\dim{(\mathcal{F}(v))}=N$ and $\dim{(\mathcal{F}(e))}=1$.  This configuration reflects the GCN architecture \cite{https://doi.org/10.48550/arxiv.1609.02907}. In GCN, each node aggregates information about its neighbours by applying a graph convolution operation.
    %The same can be obtained by setting the size of edge stalks to $1$.
    In fact, a size of edge stalks of $1$, we have no vector spaces associated to each edge and we use only neighbours' information.
    \item \textbf{\ModelName:} $\dim{(\mathcal{F}(v))}=N$ and $\dim{(\mathcal{F}(e))}=N$. Our proposed architecture features vector spaces of size $N$ for both node and edge stalks.
\end{itemize}
By fixing the size of node stalks and edge stalks to $1$ we are implicitly modifying the number of parameters. To ensure comparability between architectures,, we force all the architectures to have the same number of parameters. This is obtained by choosing the maximum number of parameters, dictated by our model ($d \times d$), and distributing the same amount of parameters in all the competitors.

Table \ref{sheaf_vs_gat_vs_gcn_table} indicates that, even with the same exact number of parameters, our proposed solution outperforms all the competitors. This underlines the expressive power of the sheaf formalism, which allows our model to effectively capture and exploit structural information within the graph.

\begin{table}[ht]
\centering
\begin{tabular}{|c @{\hspace{2em}} c|c *{3}{@{\hspace{2em}}c}|}
\hline
$\dim{(\mathcal{F}(v))}$ & $\dim{(\mathcal{F}(e))}$ & F1@10 & NDCG@10 & F1@20 & NDCG@20 \\
\hline
1 & N & 0.048 & 0.093 & 0.056 & 0.111 \\
1 & 1 & 0.045 & 0.093 & 0.055 & 0.114 \\
N & N & \textbf{0.051} & \textbf{0.105} &  \textbf{0.061} & \textbf{0.129} \\
\hline
\end{tabular}
\caption{Performance of \ModelName in terms of F1 and NDCG metrics while changing the size of node and edge stalks. The best results are obtained with the $N,N$ setting, which represents the sheaf architecture.}
\label{sheaf_vs_gat_vs_gcn_table}
\end{table}
\section{Conclusions and future work}
\label{conclusions}
Our \ModelName architecture introduces a paradigm shift by using vector spaces rather than simple vectors to represent nodes and edges within bipartite graphs of user-item interactions. Experimental results highlight \ModelName's consistent superiority over top-performing baseline models. For example, \ModelName achieves a relative improvement of 8.6\% in F1@10 and 7.6\% on NDCG@10 on MovieLens 1M, while on Yahoo! Movies it boosts NDCG@10 by 7.3\% and F1@20 by 8.7\%.

The central argument of this paper lies in the potential of novel architectures grounded in Sheaf theory. These architectures are particularly suited for contexts where relationships and their representations are complex and inherently ambiguous. Take, for instance, a user's behaviour, which is influenced by the items they interact with. Traditional vector representations may lack expressive power to capture the nuances of such behaviours, underscoring the need for a full vector space. This reasoning could extend to various application domains.

Looking forward, we aim to adapt this architecture to broader research domains. Its unique topological nature implies versatility, especially when embedding side information in related stalks. A viable avenue worth exploring is the Next Point-Of-Interest recommendation system, which could leverage historical data alongside immediate inclinations to offer personalized recommendations that enhance user experiences and satisfaction.
%\clearpage
\bibliographystyle{ACM-Reference-Format}
\bibliography{biblio}
\appendix

\end{document}